\font\bb=msbm10 at 10pt
\font\bbT=msbm10 at 7pt

\font\rms=cmr10 at 7pt

\def\0#1{\mbox{\rm#1}}
\def\1#1{\mbox{\bb#1}}
\def\2#1{\mbox{\bf#1}}
\def\3#1{{\cal #1}}
\def\4#1{\mbox{\cals#1}}
\def\5#1{\mbox{\sf#1}} %%sans serif
\def\6#1{\mbox{\frak#1}}
\def\7#1{\mbox{\fraks#1}}
\def\8#1{\mbox{\rms#1}}
\def\9#1{\mbox{\bbs#1}}

\def\BEq{\begin{equation}}
\def\EEq{\end{equation}}
\def\BEqA{\begin{eqnarray}}
\def\EEqA{\end{eqnarray}}
\def\BEn{\begin{enumerate}}
\def\EEn{\end{enumerate}}

\def\D{\Delta}

\def\f{\phi}

\def\g{\gamma}

\def\m{\mu}

\def\S{\Sigma}

\def\y{\psi}

\def\Rb{{\bf R}}

\def\Cbb{\mbox{\bb C}}
\def\Cbbs{\mbox{\bbT C}}

\def\Hbb{\mbox{\bb H}}

\def\Rbb{\mbox{\bb R}}
\def\Rbbs{\mbox{\bbT R}}

\def\tav{\hbox{

\kern-1pt\rule[0pt]{1.5pt}{.8pt}{\kern-3.3pt}

\rule[0pt]{.4pt}{5pt}{\kern-4.4pt}
\rule[5pt]{4.5pt}{.8pt}{\kern-3.45pt}
\rule[0pt]{.4pt}{5.3pt}{\kern-1pt}
}}

\def\Ac{{\cal A}}

\def\Rc{{\cal R}}

\def\Irm{{\rm I}}
\def\Irms{\mbox{\scriptsize\rm I}}

\def\Qrm{{\rm Q}}

\def\Srm{{\rm S}}
\def\Srms{{\rms S}}
\def\Srms{\mbox{\rms S}}

\def\Urm{{\rm U}}

\def\dag{\dagger}

\def\adj{{^{\dag}}}

\def\ox{\otimes}

\def\Ox{\bigotimes}

\def\pa{\partial}

\def\from{\kern-2pt\leftarrow\kern-2pt}

\def\x{\times}

\def\II{|\kern-1pt |}

\def\uar{\uparrow}

\def\Av{\mathop{\hbox{\rm Av}}\nolimits}

\def\Cliff{\mathop{\hbox{\rm Cliff}}\nolimits}

\def\End{\mathop{\rm End}\nolimits}

\def\GL{\mathop{\hbox{\rm GL}}\nolimits}

\def\Seq{\mathop{\hbox{\rm Seq}}\nolimits}

\def\Set{\mathop{\hbox{\rm Set}}\nolimits}

\def\Sib{\mathop{\hbox{\rm Sib}}\nolimits}

\def\SO{\mathop{\mbox{\rm SO}}\nolimits}

\def\Ten{\mathop{\hbox{\rm Ten}}\nolimits}

\def\BEq{\begin{equation}}

\def\EEq{\end{equation}}

\def\BEqA{\begin{eqnarray}}

\def\EEqA{\end{eqnarray}}

\def\Cliff{\mathop{\hbox{\rm Cliff}}\nolimits}

\def\End{\mathop{\hbox{\rm End}}\nolimits}

\def\Seq{\mathop{\hbox{\rm Seq}}\nolimits}

\def\Set{\mathop{\hbox{\rm Set}}\nolimits}

\def\Sib{\mathop{\hbox{\rm Sib}}\nolimits}

\def\Ten{\mathop{\hbox{\rm Ten}}\nolimits}

\def\adj{{^{\dag}}}

\def\ox{\otimes}
\def\Ox{\bigotimes}

\documentstyle{article}

\begin{document}

\title{{\bf Cliffordons}}

\author{David R. Finkelstein and Andrei A. Galiautdinov\\
\normalsize{\it School of Physics, Georgia Institute of
Technology, Atlanta, Georgia 30332-0430} \\
\\
(To appear in {\it J. Math. Phys.} {\bf 42}, August 2001)}

\maketitle

\abstract{
At higher energies
the present complex quantum theory with its unitary group
might expand into a real quantum theory with an orthogonal
group, broken by an approximate $i$ operator at lower
energies. Implementing this possibility requires a real
quantum double-valued statistics.
A Clifford statistics,
representing a swap (12) by a difference
$\gamma_1-\gamma_2$ of Clifford units,
is uniquely appropriate.
Unlike the
Maxwell-Boltzmann, Fermi-Dirac, Bose-Einstein, and para-
statistics, which are tensorial and single-valued,
and unlike anyons,
which are confined to two dimensions,
Clifford statistics
are multivalued and work for any dimensionality.
Nayak and Wilczek proposed
a Clifford statistics
for the fractional quantum Hall effect.
We apply them to toy quanta here.
A complex-Clifford example has the energy spectrum of a
system of spin-1/2 particles in an external magnetic field.
This supports the proposal that the double-valued
rotations --- spin --- seen at current energies might arise
from double-valued permutations --- swap --- to be seen
at higher energies.
Another toy with real Clifford statistics illustrates how
an effective imaginary unit $i$ can arise
naturally within a real quantum
theory. }

\pagestyle{myheadings}
\markright{{\rm D.R. Finkelstein and A.A. Galiautdinov},
Cliffordons}

\section{INTRODUCTION: QUANTIFICATION
PROCEDURES}\label{sec:GS}

Nayak and Wilczek \cite{nayakwilczek} have proposed a startling new
statistics for fractional quantum Hall effect carriers.
It has great potential for even more fundamental applications
to sub-particle structure \cite{fs}.
To learn its properties we  apply it here to some toy models.

The common statistics --- Fermi-Dirac (F-D), Bose-Einstein
(B-E) and Maxwell-Boltzmann (M-B)  --- may be regarded as
differing prescriptions
for constructing the algebra of an
ensemble of many individuals from the vector space of one
individual.
These procedures take
qualitative yes-or-no questions about an individual into
quantitative how-many questions about an ensemble of similar
individuals.
Such procedures were termed
{\it
quantification}.
Now they are sometimes called ``second
quantization,'' somewhat misleadingly.

We use a well-known operational formulation
of quantum theory.
The main point of quantum theory is that
mathematical objects may be completely describable,
since we make them up,
but physical
quanta are not.
An electron, a physical entity, is not a spinor wavefunction,
a linear operator, or any other mathematical object.
But it seems that mathematical objects can usefully represent
what we do to an electron.
Kets represent input modes
(preparation), bras represent outtake modes (registration),
operators represent intermediate operations on quantum
\cite{finkelstein96}.

Each of the usual statistics is defined
by an associated linear mapping $Q^{\dag}$ that maps any
one-body initial mode $\psi$ into a many-body creation
operator:
\BEq\label{eq:linquantification}
Q^{\dag}: V_{\Irms} \rightarrow \Ac_{\Srms}, \; \psi
\mapsto Q^{\dag}
\psi=: \hat\psi.
\EEq
Here $V_{\Irms}$ is the initial-mode vector space of the
individual $\Irm$ and $\Ac_{\Srms}=\End V_{\Srms}$ is the
operator (or endomorphism) algebra of the quantified system
$\Srm$.
The $\dag$ in $Q\adj$  reminds us that $Q\adj$
is contragredient to the initial modes $\psi$.
We write the mapping $Q\adj$ to the left of its argument
$\psi\adj$ to respect the conventional Dirac order of
cogredient and contragredient vectors in a contraction.

Dually, the final modes
$\psi^{\dag}$ of the dual space $V^{\dag}_{\Irms}$ are
mapped to annihilators in $\Ac_{\Srms}$ by the linear
operator $Q$,
\BEq
Q: V^{\dag}_{\Irms} \rightarrow
\Ac_{\Srms}, \; \psi^{\dag} \mapsto
\psi^{\dag} Q =: \hat\psi^{\dag}.
\EEq

We
call the transformation
$Q$ the {\em quantifier} for the statistics.
$Q$
and $Q^\dag$ are tensors of the type
\BEq
Q=(Q^{aB}{}_C),\quad Q^\dag =({{Q\adj_a}^C}_B),
\EEq
where $a$ indexes a basis in the one-body space $V_{\Irms}$
and $B, C$ index a basis in the many-body
space $V_{\Srms}$.

The basic creators and annihilators associated
with an { arbitrary}
basis
$\{ e_a | \, a=1,\dots, N\} \subset V_{\Irm} $ and its
reciprocal basis $\{ e^{a} | \, a =1,\dots, N\} \subset
V^\dag_{\Irm}$ are then
\BEq
\label{eq:BASICCREATORS}
Q^{\dag} e_a := \hat e_a =: Q\adj{}_a,
\EEq
and
\BEq
\label{eq:BASICANNIHILATORS}
e^{a} Q := \hat e^{a} =: Q^{a}.
\EEq
The creator and annihilator for a general
initial mode
$\psi$ are
\BEqA
Q^{\dag} (e_a \psi^a) &= &Q\adj_a \psi^a\/,\cr
(\phi\adj_{a} e^{a}) Q &=&
\phi\adj_{a} Q^{a}\/, \EEqA
respectively.

We require that quantification respects the adjpoint $\dag$.
This relates the two tensors
$Q$ and
$Q^{\dag}$:
\BEq
\label{eq:condition}
\psi^{\dag} Q = {(Q^{\dag} \psi)}^\dag . \EEq
The rightmost $\dag$ is the adjoint operation for the
quantified system.
Therefore
\BEq
\label{eq:conditionGENERATORS}
\hat{e}_a^{\dag} = M_{a b} \hat e^{b},
\EEq
with $M_{a b}$ being the metric,
the matrix of the adjoint
operation, for the individual system.

We now generalize from the common statistics.
A {\em
linear statistics} shall be defined by a linear correspondence
$Q\adj$ called the quantifier,
\BEq
\label{eq:linearquantification}
Q^{\dag}: V_{\Irms} \rightarrow \Ac_{\Srms}, \; \psi
\mapsto Q^{\dag}
\psi=: \hat\psi,
\EEq
[compare (\ref{eq:linquantification})] from one-body modes
to many-body operators,
$\dag$-algebraically generating
the algebra $\Ac_{\Srms}:=\End V_{\Srms}$ of the many-body
theory.
We further require that the quantifier $Q\adj$ induce an
isomorphism from the one-body unitary group $\Urm_{\Irms}$
into the many-body unitary group
$\Urm_{\Srms}$, as described in Sec. \ref{sec:morph}.
This is the {\em representation principle} for quantifiers.

The representation principle implies bilinear
algebraic commutation
relations discussed below.

In general $Q\adj$
does not produce a creator and $Q$ does not produce an
annihilator, as they do in the common statistics.

We construct the quantified algebra
$\Ac_{\Srms}$ from the individual space $V_{\Irms}$
in three
easy steps:

1. We form the quantum algebra
$\Ac(V_{\Irms})$, defined as the free $\dag$ algebra
generated by (the vectors of) $V_{\Irms}$.
Its elements are
all possible iterated sums and products and
$\dag$-adjoints of the vectors of $V_{\Irms}$.
We require that the  operations $(+, \x, \dag)$ of
$\Ac(V_{\Irms})$ agree with those of $V_{\Irms}$ where both
are meaningful.

2.  We construct the ideal $\Rc\subset \Ac$ of all elements of
$\Ac(V_{\Irms})$ that vanish in virtue of the statistics.
It is convenient and customary to define $\Rc$ by a set of
expressions $\Rb$, such that
the
commutation relations between elements of
$\Ac(V_{\Irms})$' have the form  $r=0$ with $r\in \Rb$.
Then $\Rc$ consists of all elements of
$\Ac(V_{\Irms})$ that vanish
in virtue of the commutation relations
and the postulates of a $\dag$-algebra.

Let $\Rb$ be closed under $\dag$.
Let $\Rc_0$ be the set of all
evaluations of all the expressions in $\Rb$
when the variable vectors
$\psi$ in these expressions assume any values $\psi \in
V_{\Irms}$.
Then $\Rc= \Ac(V_{\Irms}) \, \Rc_0\,
\Ac(V_{\Irms})$,

3. We form the quotient algebra
(actually, a residue algebra)
\BEq
\Ac_{\Srms}=  \Ac(V_{\Irms}) /\Rc,
\EEq
by identifying elements of $\Ac(V_{\Irms})$
whose differences belong to
$\Rc$.

Then $Q\adj$ maps each vector $\y\in V_{\Irms}$ into its
residue class
$\y+\Rc$.

Historically, physicists carried out
one special quantification first.
Since classically one multiplies
phase spaces when quantifying,
they assumed that quantally one multiplies
Hilbert spaces,
forming the
tensor product
\BEq
V_{\Srms}=\Ox_{p=0}^N V_{\Irms} =  V_{\Irms}^N
\EEq
of $N$ individual spaces
$V_{\Irms}$.
Then in order to improve agreement with experiment
they removed
degrees of freedom in the tensor product connected with
permutations,
reducing
$V_{\Irms}^N$ to a subspace $PV_{\Irms}^N$ invariant
under all permutations of individuals.
Here $P$ is a
projection operator characterizing the statistics.
The many-body algebra was then taken to be the algebra of
linear operators on the reduced space: $\Ac_{\Srms}=\End
P V_{\Irms}^N$.

We call a statistics built in that way
on a subspace of the tensor algebra over the one-body
initial mode space, a {\em
tensorial} statistics.
Tensorial statistics represents permutations in a single-valued way.
The common statistics are tensorial.

Linear statistics is more general than tensorial
statistics, in that the quotient algebra
$\Ac_{\Srms}=\Ac - \Rc$ defining a linear statistics need
not be the operator algebra of any subspace of the tensor
space $\Ten V_{\Irms}$ and need not be single-valued.
Commutation relations permit more general statistics
than projection
operators do.
For example, anyon statistics is linear but not tensorial.

For another example, $\Ac_{\Srms}$ may be the endomorphism algebra
of a spinor space constructed from the quadratic space
$V_{\Irms}$.
Such a statistics we call a {\em spinorial
statistics}\/.
Clifford statistics, the main topic of
this paper,
is a spinorial statistics.
Linear statistics includes both
spinorial and tensorial statistics.

The  F-D, B-E and M-B statistics are readily
presented as
tensorial statistics.
We give their quantifiers next
\cite{finkelstein96}.
We then generalize to spinorial, non-tensorial, statistics.

\section{STANDARD STATISTICS}
\label{sec:STANDARDSTATISTICS}

{\bf Maxwell-Boltzmann statistics.}
Classical
an M-B aggregate is a sequence
(up to isomorphism)
and
$Q=\Seq$, the {\it sequence}-forming quantifier.
The quantum
individual $\Irm$ has a Hilbert space
$V=V_{\Irms}$ over the field ${\Cbb}$.
The vector space for
the q sequence is the (contravariant) tensor algebra
$V_{\Srms}=\Ten V_{\Irms}$, whose product is the tensor
product $\ox$:
\BEq
V_{\Srms}=\Ten V_{\Irms}
\EEq
with the natural induced $\dag$.
The kinematic algebra $\Ac_{\Srms}$ of the sequence is the
$\dag$-algebra of endomorphisms of $\Ten V_{\Irms}$, and is
generated by $\psi \in V_{\Irms}$ subject to the generating
relations
\BEq
\hat \psi\adj \hat{\phi} = \y\adj \f. \EEq
The left-hand
side is an operator product, and the right-hand side
is the contraction of the dual vector
$\y^\dag$ with the vector $\f$, with an implicit
unit element $1\in \Ac_ {\Srm}$ as a factor.

{\bf Fermi-Dirac statistics.} Here $Q=\Set$,
the {\it
set}-forming quantifier.
The kinematic algebra for the
quantum set has defining relations
\BEqA
\hat{\y} \hat{\f} + \hat{\f}\hat \y &=&
0, \;\cr
\hat\y^{\dag}\hat \f +\hat \f \hat\y^{\dag} &=& \y\adj \f .
\EEqA for all $\y,\f\in V_{\Irms}$\/.

{\bf Bose-Einstein statistics.} Here $Q=\Sib$, the {\it
sib}-forming quantifier. The sib-generating relations are
\BEqA
\hat\y \hat\f -\hat \f \hat\y &=& 0, \;\cr \hat\y^{\dag}
\hat\f - \hat\f
\hat\y^{\dag} &=& \y\adj \f , \EEqA
for all $\y,\f\in V_{\Irms}$\/.

The individuals in each of the discussed quantifications,
by construction, have the same (isomorphic) initial spaces.
We call such individuals {\it isomorphic}.

\section{RELATION TO THE PERMUTATION GROUP}

A statistics is {\em abelian} if it represents the
permutation group $S_N$ on its $N$ individuals by an abelian
group of operators in the $N$-body mode space.

The F-D or B-E representations are not only abelian but
scalar. They represent each permutation by a number, a
projective representation of the identity operator.
One calls
entities with scalar statistics {\it indistinguishable}.
Bosons and fermions are indistinguishable.

Non-abelian statistics describe distinguishable entities.

Nayak and Wilczek \cite{wilczek, nayakwilczek} give a
spinorial statistics based on the work on nonabelions of
Read and Moore
\cite{moore90, read92}. Read and Moore use a subspace
corresponding to the degenerate ground mode of some
realistic Hamiltonian as the representation space for a
non-abelian representation of the permutation group $S_{2n}$
acting on the composite of $2n$ quasiholes in the fractional
quantum Hall effect. This statistics, Wilczek showed,
represents the permutation group on a spinor space, and
permutations by non-commuting
spin operators.
The quasiholes of Read and Moore and of Wilczek and Nayak
are distinguishable,
but their permutations leave the ground subspace invariant.

Our own interest in the statistics of distinguishable
entitities arises from a study of quantum space-time
structure \cite{fs}.
The dynamical process of any system is
composite,  it is generally believed,
composed of isomorphic
elementary actions going on all over, all the time.
The first
question that has to be answered in setting up an algebraic
quantum theory of this composite process is:
What
statistics do the elementary actions have?

The elementary processes have ordinarily,
though implicitly, been assumed to be distinguishable,
being
addressed by space-time coordinates,
and to obey Maxwell-Boltzmann statistics.
This repeats the history
of particle statistics
on the greater field of process statistics.

The Clifford statistics studied below is proposed primarily
for the elementary processes of nature.
We apply it here to toy models of particles in ordinary
space-time to familiarize ourselves with its properties.
In our construction, the representation space of the
permutation group is the { whole} (spinor) space of
the composite.
The permutation group is not assumed to be
a symmetry of the Hamiltonian or of its ground
subspace any longer.
It is used as a dynamical group, not a symmetry group.

\section{NO QUANTIFICATION WITHOUT REPRESENTATION}
\label{sec:morph}

If we have defined how, for example, one translates
individuals, this should define a way to translate the
ensemble. We shall require of a quantification that any
unitary transformation on an individual quantum entity
induces a unitary transformation on the quantified system,
defined by the quantifier.

This does not imply that, for example, the actual
time-translation of an ensemble
is carried out by translating the individuals.
This would
imply that the Hamiltonians combine additively,
without
interaction.
There is still room for arbitrary interaction.
The representation principle means only that there is a
well-defined time-translation without interaction.
This gives a physical meaning to interaction: it is the
difference between the induced time translation generator
and the actual one.

Thus we posit that an arbitrary ($\dag$-)unitary
transformation $U: V_{\Irms} \rightarrow V_{\Irms}, \, \psi
\mapsto U\psi$ of the individual ket-space
$V_{\Irms}$, also act naturally on the quantified mode
space $V_{\Srms}$ through an operator
$\hat{U}: V_{\Srms}\to V_{\Srms}$,
defining a representation of the individual unitary group.
This is the {\em representation principle}\/.

Then $U$ also acts on the algebra $\Ac_{\Srms}$ according to
\BEqA
\hat{U}:
\Ac_{\Srms} \rightarrow \Ac_{\Srms}, \, \hat\y \mapsto
\widehat{ U\y}=\hat U \hat\y \hat{U}^{-1}.
\EEqA

Every unitary transformation $U:V_{\Irms}\to V_{\Irms}$
infinitesimally different
from the identity is defined by a {\em generator} $G$: \BEq
U = 1+G\delta \theta\/,
\EEq
where $G=-{G}^{\dag}:V_{\Irms}\to V_{\Irms}$ is
anti-Hermitian and $\delta
\theta$ is an infinitesimal parameter.
The infinitesimal anti-Hermitian generators $G$ make up the
Lie algebra
$d\Urm_{\Irms}$ of the unitary group $\Urm_{\Irms}$ of the
one-body theory.

By the representation principle,
each individual generator $G$ induces a
{\em quantified generator}
$\hat G\in \Ac_{\Srms}$ of the
quantified system, defined (up to an added constant) by its
adjoint action on
$\Ac_{\Srms}$:
\BEq
\label{eq:ltfgeqa}
\hat G: \hat{\psi} \mapsto \widehat{G\y} =[\hat G,
\hat{\psi}]\/;
\EEq
and
(\ref{eq:ltfgeqa}) and (\ref{eq:quantification}) define a
representation (Lie homomorphism) $R_Q: d\Urm_{\Irms} \to
d{\Urm_{\Srms}}$ of the individual Lie algebra
$d\Urm_{\Irms}$ in the quantified Lie algebra
$d\Urm_{\Srms}$.

Since
\BEq
\label{eq:ANOTHERDEFOFOPERATOR}
G =
\sum_{a, b} e_a {{G}^a}{}_{b} e^{b}
\EEq
holds by the completeness of the basis $e_a$ and the
reciprocal basis $e^a$\/, we can express the quantified
generator $\hat G$ by \BEq
\label{eq:quantification}
\hat G := Q^{\adj} G Q =
\sum_{a , b} Q\adj_a{{G}^a}{}_{b} Q^{b}
\equiv
\sum_{a, b}\hat e_a {{G}^a}{}_{b} \hat e^{b}. \EEq

The representation principle holds for
the usual statistics (M-B, F-D, B-E)
and for the Clifford statistics discussed below.

Proposition: If $\Qrm$ is a quantifier
for a linear statistics then
\BEq
[\hat G, \, Q\adj{\psi}] = G Q\adj{\psi}
\EEq
hold for all anti-Hermitian generators $G$\/.

Proof: We have
\BEqA
\label{eq:consistency}
[\hat G, \, Q\adj{\psi}] &=
&{{G}^a}_b \, \left( \hat e_a \, \hat e^b \, Q\adj{\psi} -
Q\adj{\psi} \, \hat e_a \, \hat e^b \right) \cr
&=& {{G}^a}_b \, \left( \hat e_a \, ( e^b \psi +
(-1)^{\kappa}
\, Q\adj{\psi} \, \hat e^b) - Q\adj{\psi} \, \hat e_a \,
\hat e^b \right)
\cr &=& {{G}^a}_b \, \hat e_a \, e^b \psi \cr &=& {{G}^a}_b
\, \hat e_a \,
\psi^b \cr &=& G \, Q\adj{\psi}. \EEqA
Here $\kappa=1$ for Fermi statistics and 0 for Bose.

If $\Ac$ is any algebra, by the {\em commutator algebra}
$\D \Ac$ of $\Ac$ we mean the Lie algebra on the elements
of $\Ac$ whose product is the commutator
$[a,b]=ab-ba$ in $\Ac$.
By the commutator algebra of a quantum system $\Irm$ we
mean that of its operator algebra $\Ac_{\Irms}$.

In the usual cases of Bose and Fermi statistics, and not in
the cases of complex and real Clifford statistics discussed
below, the quantification rule (\ref{eq:quantification})
defines a Lie isomorphism, $\D
\Ac_{\Irms}\to \D \Ac_{\Srms}$, from the commutator algebra
of the individual
to that of the quantified system.
Namely,
if $H$ and
$P$ are two (arbitrary) operators acting on the
one-body ket-space, then
\BEq
\label{eq:Lie}
\widehat{[H, \, P]} = [\hat H, \, \hat P].
\EEq
Explicitly,
\BEqA
[\hat H, \, \hat P] &= & \hat H \hat P - \hat P \hat H \cr
&= & \hat e_r{{H}^r}_s\hat e^s \;
\hat e_t{{P}^t}_u\hat e^u
-\hat e_t{{P}^t}_u\hat e^u \;
\hat e_r{{H}^r}_s\hat e^s \cr
&= & {{H}^r}_s {{P}^t}_u
(\hat e_r \hat e^s \hat e_t \hat e^u
-\hat e_t \hat e^u \hat e_r \hat e^s) \cr &= & {{H}^r}_s
{{P}^t}_u (\hat e_r (\delta_t^s \pm \hat e_t \hat e^s) \hat
e^u -\hat e_t \hat e^u
\hat e_r \hat e^s) \cr
&= & {{H}^r}_s {{P}^t}_u
(\hat e_r \delta_t^s \hat e^u \pm \hat e_r \hat e_t \hat
e^s \hat e^u -\hat e_t \hat e^u \hat e_r \hat e^s) \cr
&= & {{H}^r}_s {{P}^t}_u
(\hat e_r \delta_t^s \hat e^u \pm \hat e_t \hat e_r \hat
e^u \hat e^s -\hat e_t \hat e^u \hat e_r \hat e^s) \cr
&= & {{H}^r}_s {{P}^t}_u
(\hat e_r \delta_t^s \hat e^u \pm \hat e_t (\mp \delta^u_r
\pm \hat e^u
\hat e_r) \hat e^s -\hat e_t \hat e^u \hat e_r \hat e^s) \cr
&= & {{H}^r}_s {{P}^t}_u
(\hat e_r \delta_t^s \hat e^u - \hat e_t \delta^u_r \hat
e^s) \cr &= & \hat e_r ( {{H}^r}_t {{P}^t}_u -
{{P}^r}_t {{H}^t}_u) \hat e^u \cr
&= & \widehat{[H, \, P]}.
\EEqA

This implies that for B-E and F-D statistics,
the
quantification rule (\ref{eq:quantification})
can be extended from the
unitary operators and their anti-Hermitian generators to the whole
operator algebra of the quantified system.

\section{CLIFFORD QUANTIFICATION}
\label{sec:CS}

Now let the one-body mode space $V_{\Irms} =
{\Rbb}^{N_+,N_-}=N_+\Rbb
\oplus N_-\Rbb$
be a real quadratic space of dimension $N=N_+ + N_-$ and
signature
$N_+-N_-$.
Denote the symmetric metric form of $V_{\Irms}$
by $g=(g_{ab}):=(e_a^\dag e_b)$.
We do not assume that $g$ is positive-definite.

We define {\em Clifford quantification}
(\ref{eq:linearquantification}) by

(1) the Clifford-like generating relations
\BEqA
\label{eq:spinorialquantification}
\hat\y \hat\f+\hat\f \hat\y &=&  \, \frac{\zeta}{2} \,
\y\adj \f
\EEqA
for all $\f, \y\in V_{\Irms}$,
where $\zeta$ is a $\pm$ sign that can have either value.

(2) the Hermiticity condition (\ref{eq:condition})
\BEq
\label{eq:conditionGENERATORSCLIFFORD}
\hat{e}_a^{\dag} = g_{a b} \hat e^{b},
\EEq

(3) a rule for raising and lowering indices
\BEq
\label{eq:conditionGENERATORSCLIFFORD}
\hat{e}_a:= \zeta' \, g_{a b} \hat e^{b},
\EEq
where $\zeta'$ is another $\pm$ sign,
and

(4) the definition (\ref{eq:quantification}) to quantify
one-body generators.

Here $\zeta = \pm 1$ covers
the two different conventions used in the
literature.
In Sec. \ref{sec:QUANTIFYINGOBSERVABLES} we will see that
$\zeta=\zeta'$, and that
$\zeta= \zeta' = +1$ and $\zeta=\zeta' = -1$ are both
allowed physically at the present theoretical stage
of development.
They lead to two
different real quantifications, with either Hermitian or
anti-Hermitian Clifford units.

For the quantified
basis elements of $V_{\Irm}$
(\ref{eq:spinorialquantification}) leads to
\BEqA
\hat e_a \hat e_b + \hat e_b \hat e_a &=&
\frac{\zeta}{2} \, g_{ab}.
\EEqA

The $\y$'s, which are assigned grade 1
and taken to be either Hermitian or anti-Hermitian, generate
a graded
$\dag$-algebra that we call the
{\it free Clifford $\dag$-algebra}
associated with
${\Rbb}^{N_+,N_-}$ and write as
$\Cliff (N_+, N_-)$ $\equiv\Cliff(N_\pm)$.
$\Cliff(N_\pm)$ contains a double-valued (or projective)
representation of the permutation group $S_N$.

Clifford statistics assembles cliffordons
individually described by
vectors into a composite described by spinors,
which we call a {\em squadron}\/.
We intend the -on suffix to remind us that unlike the common
statistics the Clifford statistics has no classical
correspondent.

A cliffordon is a hypothetical quantum-physical entity,
like an electron,
not to be confused with a mathematical object like a spinor
or an operator.
We cannot describe a cliffordon completely,
but we represent our actions on a squadron of cliffordons
adequately by operators in a Clifford algebra of operators.
One
encounters cliffordons only in permuting them,
never in
creating or annihilating them as individuals.

In assuming a real vector space of quantum modes
instead of a complex one,
we give up $i$-invariance but retain quantum superposition
$a\psi + b \phi$ with real coefficients.
Our theory is non-linear from the complex point of
view.
Others considered non-linear quantum theories,
but gave up real superposition as well as $i$-invariance
\cite{weinberg},\cite{brody}.
We are not {\it that} non-linear.

\section{QUANTIFYING OBSERVABLES}
\label{sec:QUANTIFYINGOBSERVABLES}

In the usual statistics, the quantifier $Q$ can be usefully
extended from the Lie algebra of the individual to the
commutator algebra of the individual;
that is, from
anti-Hermitian operators to all operators.
This is not the case for Clifford quantification.
There the quantification of any symmetric operator is a
scalar, in virtue of Clifford's law,
and so the commutator of any two operators is just the
commutator of their antisymmetric parts. A straightforward
calculation shows that
\BEqA
[\hat H, \, \hat P] & = &
\hat H \hat P - \hat P \hat H \nonumber \\
&=& \zeta \, \zeta' \, \left( \frac{1}{2} \widehat{[H,
\, P]} +
\frac{1}{4} \left( \widehat{[P, \, H\adj]} +
\widehat{[P\adj, \, H]} \right) \right).
\EEqA

The three simplest cases are
\begin{enumerate}

\item $H=H\adj$, $H'=H'\adj$ $\Longrightarrow$ $[\hat H,
\hat H'] = 0$;

\item $H=H\adj$, $G_1=-G_1\adj$ $\Longrightarrow$ $[\hat H,
\, \hat G_1] = 0$;

\item $G=-G\adj$, $G'=-G'\adj$ $\Longrightarrow$ $[\hat G,
\, \hat G'] = \zeta \, \zeta' \,
\widehat{[G, \, G']}$.

\end{enumerate}

  Thus Clifford quantification respects the commutation relations
for anti-Hermitian generators
if and only if $\zeta = \zeta' = +1$ or $\zeta= \zeta' =
-1$; but not for Hermitian
observables, contrary to the Bose and Fermi quantifications,
which respect both.

\section{NAYAK-WILCZEK STATISTICS}
\label{sec:cwstatistics}

The {\em complex} graded algebra generated by the $\y$'s
with the relations (\ref{eq:spinorialquantification})
is called the {\em complex Clifford algebra}
$\Cliff_{\Cbbs}(N)$ over
${\Rbb}^{N_+,N_-}$. It is isomorphic
to the full complex matrix algebra
${\Cbb (2^n)\otimes \Cbb (2^n)}$ for even $N=2n$, and to
the direct sum
${\Cbb(2^n)\otimes \Cbb(2^n)} \oplus {\Cbb(2^n)\otimes
\Cbb(2^n)}$ for odd
$N=2n+1$.
We regard $\Cliff_{\Cbbs}(N)$ as the kinematic algebra of
the complex Clifford composite.
As a vector space, it has dimension $2^N$.

Schur \cite{schur} used complex spinors and complex Clifford
algebra to represent permutations some years before Cartan
used them to represent rotations.
There is a fairly widespread view that spinors may be more
fundamental than vectors,
since vectors may be expressed as bilinear
combinations of spinors.
One of us took this direction in much of his
work.
Clifford statistics support the opposite view.
There a vector describes an individual, a spinor an aggregate.
Wilczek and Zee \cite{Wilczek82} seem to have been the
first to recognize
that spinors represent composites in a physical
context,
although this is implicit
in the Chevalley construction of spinors
within a Grassmann algebra.

For dimension $N=3$
spinors have as many parameters as vectors, but
for higher $N$ the number of components of the spinors
associated with $\Cliff(N_\pm)$
grows exponentially with $N$.
The physical relevance of this
irreducible double-valued (or projective) representation of
the permutation group
$S_{N}$
was recognized by Nayak and Wilczek
\cite{nayakwilczek,wilczek}
in a theory of the fractional quantum Hall effect.
We call the complex statistics
based on
$\Cliff_{\Cbbs}(N)$
the {\it Nayak-Wilczek} or N-W statistics.

Clifford statistics, unlike the more familiar particle
statistics \cite{berezin, feynman, negele},
provides no
creators or annihilators.
With each individual mode $e_a$ of
the quantified system they associate a Clifford unit
$\gamma_a = 2Q\adj_a$\/.

We may represent any swap
(transposition of two cliffordons, say 1 and 2)
by the difference of the corresponding Clifford
units
\BEq
t_{12}:=\frac{1}{\sqrt{2}}(\gamma_1 - \gamma_2).
\EEq
and represent an arbitrary permutation, which is a product of
elementary swaps,
by the product of their representations.
That is, as direct computation shows,
this defines a projective homomorphism
from $\0S_N$ into the Clifford algebra
generated by the $\gamma_k$\/.

By definition, the number $N$
of cliffordons in a squadron is the dimensionality of the
individual initial mode space $V_{\Irms}$.
$N$ is conserved
rather trivially, commuting with every Clifford element.
We
can change this number only
by varying the dimensionality of the one-body space.
In one use of the theory,
we can do this, for example,
by changing the space-time 4-volume of the corresponding
experimental region.
Because our theory does not use creation and annihilation
operators, an initial action on the squadron represented by a
spinor $\xi$ should be viewed as some kind of spontaneous
transition condensation into a coherent mode, analogous to
the transition from the superconducting to the many-vertex
mode in a type-II superconductor. The initial mode of a set
or sib of (F-D or B-E) quanta can be regarded as a result of
possibly entangled creation operations. That of a squadron of
cliffordons cannot.

As with (\ref{eq:consistency}),
let us verify that definition (\ref{eq:quantification}) is
consistent in the Clifford case:
\BEqA
\label{eq:consistencyCW}
[\hat G, \, Q\adj \psi]&=
&{{G}^a}_b \, \left( \hat e_a \, \hat e^b \, Q\adj \psi -
Q\adj \psi \,
\hat e_a \, \hat e^b \right) \cr &=& \frac{1}{2} {{G}^a}_b
\, \left( \hat e_a \psi^b + \psi_a \hat e^b \right) \cr &=&
G \, Q\adj \psi.
\EEqA
This shows that $Q\adj \psi$ transforms correctly under the
infinitesimal unitary transformation of ${\Rbb}^{N_+,N_-}$
({\it cf.}
\cite{dirac74}).

\section{BREAKING $i$ INVARIANCE}

Thus we cannot construct useful Hermitian variables of a
squadron by applying the quantifier to the Hermitian
variables of the individual cliffordon.

This is closely related to fact that the real initial mode
space
$\Rbb^{N_\pm}$ of a cliffordon has no special operator
to replace the imaginary unit $i$ of the standard
complex quantum theory. The fundamental task of the
imaginary element $i$ in the algebra of complex quantum
physics is precisely to relate conserved Hermitian
observables $H$ and anti-Hermitian generators $G$
by
\BEq
\label{eq:GHI}
H=-i\hbar G.
\EEq
To perform this function exactly, the operator $i$ must
commute exactly with all observables.

The central operators $x$ and $p$ of classical mechanics
are contractions of noncentral operators $\breve{x}$
and $\breve{p}=-i\hbar \pa/\pa \breve{x}$ \cite{fs}.
In the limit of
large numbers of individuals organized coherently into
suitable condensate modes, the expanded operators of the
quantum theory contract into the central operators of the
classical theory.
Condensations produce nearly commutative variables.

Likewise we expect the central
operator $i$ to be a contraction of a non-central operator
$\breve{i}$ similarly resulting from a condensation
in a limit of large numbers.
In the simpler expanded theory,
$\breve{i}$, the correspondent of $i$, is not central.

One clue to the nature of $\breve{i}$ and the locus of its
condensation is how the
operator $i$ behaves when we combine separate systems. Since
infinitesimal generators $G, G',
\dots $ combine by addition, the imaginaries $i, i', \dots
$ of different individuals must combine by identification
\BEq
i=i'=\dots
\EEq
for (\ref{eq:GHI}) to hold exactly,
and nearly so for (\ref{eq:GHI}) to hold nearly. The only
other variables in present physics that combine by
identification in this way are the time
$t$ of clasical mechanics and the space-time coordinates
$x^\m$ of field theories. All systems in an ensemble must
have about the same $i$, just as all particles have about
the same $t$ in the usual instant-based formulation, and all
fields have about the same space-time variables $x^{\m}$ in
field theory. We identify the variables $t$ and $x^\m$ for
different systems because they are set by the experimenter,
not the system.  This suggests that the
experimenter, or more generally the environment of the
system, mainly defines the operator
$i$.
The central operators $x, p$ characterize a small system
that results from the condensation of many particles.
The central operator $i$ must result from a condensation in
the environment; we take this to be the same condensation
that forms the vacuum and the spatiotemporal structure
represented by the variables
$x^\m$ of the
standard model.

The existence of this contracted $i$
ensures that at least approximately,
every Lie commutation relation
between dimensionless anti-Hermitian generators $A, B, C$
of the  standard complex quantum theory,
\BEqA
\label{eq:relation}
[A, \, B]=C,
\EEqA
corresponds to a commutation relation between Hermitian
variables $-i\hbar A$, $-i\hbar B$, $-i\hbar C$:
\BEqA
[-i\hbar A, \, -i\hbar B] = -i\hbar(-i\hbar C) . \EEqA
It also tells us that this correspondence is not exact in
nature.

St\"{u}ckelberg \cite{stuckelberg} reformulated
complex quantum mechanics in the real Hilbert space
${\Rbb}^{2N}$ of twice as many dimensions by assuming a
special real antisymmetric operator $J: \Rbb^{2N}\to
\Rbb^{2N}$ commuting with all of the variables of the
system.

A real $\dag$ or Hilbert space
has no such operator.
For example, in ${\Rbb}^{2}$ the operator
\BEq
E:=
\left[\matrix{\varepsilon_1 & 0 \cr
0& \varepsilon_2}\right]
\EEq
is a symmetric operator
with an obvious spectral decomposition representing,
according to the usual interpretation, two selection
operations performed on the system, and cannot be written in
the form $G=-J \hbar E$ relating it to some antisymmetric
generator $G$ for any real antisymmetric $J$ commuting with
$E$.

       On the other hand, if we restrict ourselves to observable
operators of the form
\BEq
E':=
\left[\matrix{\varepsilon & 0 \cr
0& \varepsilon}\right],
\EEq
we can use the operator $J$,
\BEq
\label{eq:J}
J:=
\left[\matrix{0 & 1 \cr
-1& 0}\right],
\EEq
to restore the usual connection between symmetry
transformations and corresponding observables.
This restriction can be generalized to any even
number of dimensions
\cite{stuckelberg}.

\section{BREAKDOWN OF THE EXPECTATION VALUE FORMULA}

For a system described in terms
of a general real Hilbert space there is no simple relation
of the form $G=\frac{i}{\hbar}H$ between the symmetry
generators and the observables: the usual
notions of Hamiltonian and momentum are meaningless in that
case. This amplifies
our earlier observation that Clifford quantification $A\to
\hat A$ respects the Lie commutation relations among
anti-Hermitian generators, not Hermitian observables.

       Operationally, this means that selective acts of individual
and quantified cliffordons use essentially different sets of
filters. This is not the case for complex quantum mechanics
and the usual statistics. There some important filters for
the composite are simply assemblies of filters for the
individuals.

Again, in the complex case
the expectation value formula for an assembly
\BEq
\Av X = \psi \adj X \psi/\psi\adj \psi
\EEq
is a consequence of the eigenvalue principle for
individuals, rather than an independent assumption
\cite{finkelstein63, finkelstein96}. The argument presented
in \cite{finkelstein63, finkelstein96} assumes that the
individuals over which the average is taken combine with
Maxwell-Boltzmann statistics. For highly excited systems
this is a good approximation even if the individuals have
F-D or B-E or other tensorial statistics.
It is not necessarily a good approximation for cliffordons,
which have spinorial, not tensorial, statistics.

\section{SPIN-1/2 COMPLEX CLIFFORD MODEL}
\label{sec:morerealistic}

In this section we present a simplest possible model of a
complex Clifford composite. The resulting many-body energy
spectrum is isomorphic to that of a sequence of spin-1/2
particles in an external magnetic field.

Recall that in the usual complex quantum theory the
Hamiltonian is related to the infinitesimal time-translation
generator $G=-G\adj$ by $G=iH$. Quantifying $H$ gives the
many-body Hamiltonian. In the framework of spinorial
statistics, as discussed above, this does not work, and
quantification in principle applies to the anti-Hermitian
time-translation generator $G$, not to the Hermitian
operator $H$. Our task now is to choose a particular
generator and to study its quantified properties.

We assume an  even-dimensional real initial-mode space
$V_{\Irms}=\Rbb^{2n}$ for the quantum individual, and
consider the
dynamics with the
simplest non-trivial
time-translation generator
\begin{eqnarray}
\label{eq:G}
G:= \varepsilon
\left[\matrix{{\bf 0}_n & {\bf 1}_n \cr
-{\bf 1}_n& \hphantom{-}{\bf 0}_n}\right] \end{eqnarray}
where $\varepsilon$ is a constant energy coefficient.

The quantified time-translation generator $\hat G$ then has
the form
\begin{eqnarray}
\label{eq:hat G}
\hat{G}&:= &\sum_{l, j}^N
\hat e_l G^l{}_j \hat e^j
\cr
&= & -\varepsilon \sum_{k=1}^{n}
(\hat e_{k+n}\hat e^k -
\hat e_k\hat e^{k+n}) \cr
&= & +\varepsilon \sum_{k=1}^{n}
(\hat e_{k+n}\hat e_k -
\hat e_k\hat e_{k+n}) \cr
&= & 2 \varepsilon \sum_{k=1}^{n}\hat e_{k+n}\hat e_k \cr
&\equiv &
\frac{1}{2} \varepsilon \sum_{k=1}^{n} \gamma_{k+n}
\gamma_k. \end{eqnarray}

By Stone's theorem, the generator $\hat G$ of time
translation in the spinor space of the complex Clifford
composite of $N=2n$ individuals can be factored
into a Hermitian $H^{(N)}$ and an imaginary unit $i$ that
commutes strongly with $H^{(N)}$:
\BEq
\hat G = i H^{(N)}.
\EEq
We suppose that $H^{(N)}$ corresponds to the Hamiltonian
and seek its spectrum.

We note that by (\ref{eq:hat G}), $\hat G$ is a sum of $n$
commuting anti-Hermitian algebraically independent operators
$\gamma_{k+n} \gamma_k, k=1,2,...,n$,
$(\gamma_{k+n} \gamma_k)^{ \dagger }=-\gamma_{k+n}
\gamma_k$,
$(\gamma_{k+n} \gamma_k)^2=-1^{(N)}$.

We use the well-known $2^n \times 2^n$
complex matrix representation of the
$\gamma$-matrices  of the complex universal Clifford
algebra associated with the real quadratic space
$\Rbb^{2n}$  (Brauer and Weyl \cite{brauer}):
\begin{eqnarray}
\gamma_{2j-1} = \sigma_3 \otimes \cdots \otimes \sigma_3
\otimes
\sigma_1 \otimes {\bf 1} \otimes \cdots \otimes {\bf 1},
\cr \gamma_{2j} =
\sigma_3 \otimes \cdots \otimes \sigma_3 \otimes 
\sigma_2 \otimes {\bf 1} \otimes \cdots \otimes {\bf 1}, \cr
j = 1, \, 2, \, 3,..., \, n,
\end{eqnarray}
where $ \sigma_1$, $ \sigma_2$ occur in the $j$-th
position, the product involves $n$ factors, and $\sigma_1$,
$\sigma_2$, $\sigma_3$ are the Pauli matrices.
The representation of the corresponding permutation group
$S_{2n}$ is  reducible.
We can simultaneously
diagonalize the $2^n\times2^n$ matrices representing the
commuting operators $\gamma_{k+n} \gamma_k$, and use their
eigenvalues, $\pm i$, to find the spectrum $\lambda$ of
$\hat G$, and consequently of $H^{(N)}$.

       A simple calculation shows that the spectrum of $\hat G$
consists of the eigenvalues
\BEq
\lambda_k = \frac{1}{2} \varepsilon (n - 2k)i,\quad k=0, \,
1, \, 2,\dots,
\, n\/,
\EEq
with multiplicity
\BEq
\m_k= C^n_k:=\frac{n!}{k!(n-k)!} \/.
\EEq

The spectrum of eigenvalues of the Hamiltonian $H^{(N)}$
then consists of $n+1$
energy levels
\BEq
\label{eq:spectrum H}
E_k =  \frac{1}{2}(n - 2k) \varepsilon ,
\EEq
with degeneracy $\m_k$.
Thus $E_k$ ranges over the interval
\BEq
-\frac{1}{4} N \varepsilon < E < \frac{1}{4} N \varepsilon,
\EEq
in steps of $\varepsilon$, with the given degeneracies.

Thus the spectrum of the structureless
$N$-body complex Clifford composite
is the same as that of a system of $N$ spin-1/2
Maxwell-Boltzmann particles of magnetic moment $\m$ in a
magnetic field {\bf H}, with the identification \BEq
\frac{1}{4} \varepsilon = \mu H\/.
\EEq

Even though we started with such a simple one-body
time-translation generator as (\ref{eq:G}),  the
spectrum of the resulting many-body Hamiltonian possesses
some complexity, reflecting the fact
that the units in the composite are distinguishable, and
their swaps generate the
dynamical variables of the system.

This spin-1/2 model
does not tell us how to swap two
Clifford units experimentally.
Like the phonon model of the harmonic oscillator,
the statistics of the
individual quanta enters the picture only through the
commutation relations among the fundamental operators of the
theory.

\section{REAL CLIFFORD STATISTICS}
\label{sec:cliffordstatistics}

Real Clifford quantification establishes a morphism
(\ref{eq:quantification}) from the Lie algebra of the
individual into that of the composite.
The proof for real Clifford statistics parallels that for
the complex Clifford case closely.

According to the Periodic Table of the Spinors
\cite{budinich, lounesto, porteous, snygg},
the free (or universal) Clifford
algebra
$\Cliff_{\Rbbs}(N_+, N_-)$ is algebra-isomorphic to the
endomorphism algebra of a module $\S(N_+, N_-)$ over a
ring ${\cal R}(N_+, N_-)$.
We give the table to simplify reference to it
\cite{budinich} (here $\zeta = - 1$):

\BEq
\label{eq:PERIODIC}
\begin{array}{l|cccccccccccccc}
&N_- &  0  &  1   &  2  &  3  &  4  &  5  &  6  &  7
&\dots&
\\ \hline
N_+&\\

0 && \1R & \1R_2 & 2\1R & 2\1C & 2\1H & 2\1H_2 & 4\1H &
8\1C &\dots
\\
1
&  &\1C & 2\1R & 2\1R_2 & 4\1R& 4\1C & 4\1H & 4\1H_2 &
8\1H&\dots
\\
2
& & \1H & \1C_2 & 4\1R & 4\1R_2 & 8\1R& 8\1C & 8\1H &
8\1H_2 &\dots
\\
3
& &\1H_2& 2\1H & 4\1C & 8\1R & 8\1R_2 & 16\1R& 16\1C &
16\1H &\dots
\\
4 & &
2\1H &2\1H_2& 4\1H & 8\1C & 16\1R & 16\1R_2 & 32\1R&
32\1C &\dots
\\
5
& &
4\1C &
4\1H &4\1H_2& 8\1H & 16\1C & 32\1R & 32\1R_2 & 64\1R &\dots
\\
6
        & & 8\1R& 8\1C &
8\1H &8\1H_2& 16\1H & 32\1C & 64\1R & 64\1R_2 &\dots
\\
7
        &  &8\1R_2
        & 16\1R& 16\1C &
16\1H &16\1H_{2}& 32\1H & 64\1C & 128\1R &\dots
\\
\vdots & &\vdots &\vdots &\vdots &\vdots &\vdots &\vdots
&\vdots &\vdots
\\
\end{array}
\EEq

It shows that the ring of coefficients
${\cal R}(N_+, N_-)$ varies periodically with period 8 in
each of the dimensionalities
$N_{+}$ and
$N_-$ of
$V_{\Irms}$, and is a function of signature $N_{+}-N_-$
alone. In the first cycle, $N_+-N_-=0, 1, \dots, 7$\/, and
${\cal R}=\Rbb,\;
\Cbb,\;
\Hbb, \;
\Hbb\oplus\Hbb,\;
\Hbb,\;
\Cbb,\;
\Rbb, \;
\Rbb\oplus \Rbb$, respectively.
Then the
cycle repeats ad infinitum.

In our application the module $\S(N_+, N_-)$, the spinor
space supporting
$\Cliff_{\Rbbs}(N_+, N_-)$,  serves as the initial mode
space of a squadron of $N$ real cliffordons.
${\cal R}(N_\pm)$ we call the {\em spinor
coefficient ring}
for $\Cliff_{\Rbbs}(N_+, N_-)$.

\section{PERMUTATIONS}

In the standard statistics there is a natural
way to represent permutations of individuals in the
$N$-body composite.
Each $N$-body ket is constructed
by successive action of $N$ creation operators on the
special vacuum mode.
Any permutation of individuals can be achieved
by permuting these creation operators in the product.
The identity and alternative representations of the
permutation group
$S_N$ in the B-E and F-D cases then follow from the defining
relations of Sec. \ref{sec:STANDARDSTATISTICS}.

In the case of Clifford statistics, some things are
different. There is still an operator associated with each
cliffordon; now it is a Clifford unit.
Permutations of cliffordons are still
represented by operators on a many-body $\dag$ space.
But
the mode space on which these operators act is now a spinor
space,
and its  basis vectors are not constructed by
creation operators acting on a special ``vacuum'' ket.

The Clifford representation of the permutation group that
we have employed is reducible into two irreducible Schur
representations. It is a bit easier to write than Schur's
because our individual operators $\gamma_i$
anticommute exactly, corresponding to exactly orthogonal
directions in the one-body mode space,
like the generators of Dirac's Clifford algebra.
In Schur's irreducible representation (slightly simplified)
these operators are replaced by their projections
normal to the principle diagonal direction $n:= \sum
\gamma_i/ \sqrt N$,  which is invariant under all
permutations. The corresponding angles
are those subtended by the edges of a regular simplex of $N$
vertices in $N-1$ dimensions as seen from the center.
These angles are all determined by
\BEq
\cos^2\theta={1\over N-1}\/.
\EEq
They differ from
$\pi/2$ by an angle that vanishes for large $N$ like
$1/N$.

\section{EMERGENCE OF A QUANTUM $i$}

The Periodic Table of the Spinors
(section {\ref{sec:cliffordstatistics}}) suggests another origin
for the complex $i$ of quantum theory, and one that is not approximately
central but exactly central.
Some Clifford algebras
$\Cliff_{\Rbbs}(N_+, N_-)$ have the spinor coefficient ring
$\Cbb$, containing an element $i$.
Multiplication by this $i$
then represents an  operator in the center of the
Clifford algebra, which we designate also by $i$.
We may use $i$-multiplication to represent the top element
$\g^{\uar}$ whenever $\g^{\uar}$  is central and has square $-1$.
This $i\in \Cliff_{\Rbbs}(N_\pm)$ corresponds
to the $i$ of complex quantum theory.

$\Cliff_{\Rbbs}(1, 0)$ contains such an $i$ but is
commutative. According to the Periodic Table (with the
choice of $\zeta = -1$), the smallest non-commmutative
Clifford algebras of  Euclidean signature with complex
spinor coefficients are  $\Cliff_{\Rbbs}(0,3)$ with negative
Euclidean signature, and $\Cliff_{\Rbbs}(5,0)$ with positive
Euclidean signature. Triads or pentads of such cliffordons
could underlie the physical ``elementary'' particles, giving
rise to complex quantum mechanics within the real.
We consider these two cases in turn.

$\Cliff_{\Rbbs}(0, 3)=\Cbb (2)$ has the familiar Pauli representation
$\gamma_1 := i \, \sigma_1, \; \gamma_2 := i
\, \sigma_2 , \; \gamma_3 := i \, \sigma_3$
with $\zeta=-1$\/.
We choose a particular one-cliffordon dynamics of
the form
\BEq
G := \left[\matrix{ 0 & V & 0  \cr -
V & 0 & \varepsilon \cr
      0& -\varepsilon &0}\right].
\EEq
Quantification (\ref{eq:quantification}) of $G$ gives
\BEq
\hat G = i \; H^{(-3)}
\EEq
with the Hamiltonian
\BEq
H^{(-3)} = \frac{1}{2}
\left[\matrix{ V & \varepsilon \cr
\varepsilon &-V}\right].
\EEq
This is also the Hamiltonian for a generic two-level
quantum-mechanical system (with the energy separation
$\varepsilon$) in an external potential field $V$, like
the ammonia molecule in a static electric field
discussed in \cite{FEYNMANLECTURES}.

$\Cliff_{\Rbbs}(5, 0)=\Cbb (4)$ has the matrix representation
$\gamma_1 := i \, \sigma_1 \otimes {\bf 1}, \; \gamma_2 := i
\, \sigma_2 \otimes {\bf 1}, \; \gamma_3 := i \, \sigma_3
\otimes \sigma_1, \; \gamma_4 := i \, \sigma_3 \otimes
\sigma_2, \; \gamma_5 := i \, \sigma_3 \otimes \sigma_3$,
again with $\zeta = - 1$. Its top Clifford unit is
$\gamma^{\uar}:= \, \prod_k \gamma_k\/= \gamma^{\uar\dag}=
\gamma^{\uar -1}$ with eigenvalues $\pm 1$\/.
We choose a specimen dynamics (for the
individual cliffordon)  in
the form
\BEq
G := \left[\matrix{ 0 & V & 0 & 0 & 0 \cr - V & 0 & 0 & 0 & 0
\cr 0&0&0&0&0 \cr 0&0&0&0&0 \cr 0&0&0&0&0 }\right].
\EEq
Quantification (\ref{eq:quantification}) of $G$ gives
\BEq
\hat G = i \; H^{(5)}
\EEq
with the Hamiltonian
\BEq
H^{(5)} = \frac{1}{2} V
\left[\matrix{ 1 & 0 & 0 & 0 \cr
0 & 1 & 0 & 0 \cr
0&0&-1&0 \cr 0&0&0&-1}\right].
\EEq

The two examples considered above show how a squadron of
several real cliffordons can obey a truly complex quantum
theory.

\section{FERMI AND CLIFFORD STATISTICS}

The F-D algebra of creators and
annihilators is a special case  of a Clifford algebra
over a quadratic space with neutral quadratic
form, called the quantum algebra by Saller \cite{saller}
and the mother algebra by Doran {\it et al.} \cite{doran}.
Is F-D statistics ever a special case of
Clifford statistics?
Specifically, are their $\dag$-algebras ever isomorphic?

  From the $N$ annihilators $a_k$ of the complex F-D statistics
we can form a
sequence of anti-commuting hermitian square roots of unity
\BEq
i_k={a_k+a_k\adj}, \quad i_{k+N}={a_k-a_k\adj\over i}
\EEq
Moreover, the complex $\dag$-algebra that these generate is a Clifford
$\dag$-algebra
$\Cliff(2N,0)$.
The transformation from the F-D generators to the Clifford is invertible.
Therefore complex F-D statistics
and complex Clifford statistics have isomorphic $\dag$-algebras.

The graded $\dag$-algebras are obviously not isomorphic.
The two grade operators do not even commute.

The question is more complicated for the real Fermi and Clifford
quantifications.
We follow Doran
{\it et al.} \cite{doran}, among others.

In the real F-D formulation we begin with a real
one-fermion $n$-dimensional space
$F\cong n\1R $ with no metric or adjoint.
The F-D quantified algebra $\Ac$ has the bilinear associative
product defined by the F-D relations
\BEqA
\label{eq:FD}
f_i f_j + f_j f_i &=& 0, \nonumber \\
f_i f^j + f^j f_i &=& \delta_i^j,
\EEqA
and the adjoint defined by
\BEq
f_i\adj := f^i.
\EEq
The $f_i$ are
creation and $f^j$ are annihilation operators.

To present $\Ac$ as a Clifford algebra
we form the direct sum
\BEq
W=F \oplus F\adj.
\EEq
In a basis $\{ f_i, \,  f^i\}_{i=1}^n $
adapted to this direct sum, we define the following
$\GL(V)$-invariant metric for
$W$:
\BEq
g \sim \left[\matrix{ 0 & 1/2 \cr 1/2 & 0 }\right],
\EEq
corresponding to
\BEq
\label{eq:QUADFORM1}
f_i \cdot f_j =0, \quad f^i \cdot f^j =0, \quad f^i \cdot f_j
= \frac{1}{2} \delta^{i}_{j}.
\EEq

Since $F$ supports a quantum theory it too has a quadratic form
$*: F\ox F\to\1R$, which we assume to be Euclidean.
We did not use $*$ in the construction of $\Ac$ and $g$.

We quantify this fermion by a mapping
$Q\adj: W\to \Ac$
into the $\dag$-algebra of the composite.
For brevity we write $f_i$ for $Q\adj f_i$
as is also customary.

The quantification $Q$
has the representation property.
In the F-D case this means that $Q$
represents the orthogonal
group $\SO(F, *)$ in $\Ac$;
in fact it represents
the larger group $\GL(F)$,
for $*$ has not entered into the
definition of $Q$.

The basis $\{\gamma_i, \, \tilde{\gamma}_i\}_{i=1}^n$
defined by
\BEq
\gamma_i := f_i +f^i, \quad
\tilde{\gamma}_i :=f_i - f^i,
\EEq
gives the metric $g$ of $W$ the diagonal form
\BEq
g \sim \left[\matrix{ 1 & 0 \cr 0 & -1 }\right],
\EEq
corresponding to
\BEq
\gamma_i \cdot \gamma_j =1, \quad \tilde{\gamma}_i \cdot
\tilde{\gamma}_j = -1,
\quad
\tilde{\gamma}_i \cdot \gamma_j = 0.
\EEq
That is, $W=E\oplus \tilde{E}$ is a neutral
quadratic space, with Eucidean subspace $E$ and
anti-Euclidean subspace $\tilde{E}$.

The $\gamma$'s obey
\BEqA
\label{eq:CLIFFORDW}
\gamma_i\gamma_j+\gamma_j\gamma_i=+2\delta_{ij}, \nonumber \\
\tilde{\gamma}_i\tilde{\gamma}_j
+ \tilde{\gamma}_j\tilde{\gamma}_i=-2\delta_{ij},
\nonumber \\
\tilde{\gamma}_i \gamma_j+\gamma_j\tilde{\gamma}_i=0.
\EEqA
Therefore the F-D algebra (\ref{eq:FD})
is isomorphic to a real Clifford algebra
$\Cliff(W,\dag)=\Cliff(E\oplus
\tilde{E})$.

Are the Clifford and F-D
$\dag$-algebras also isomorphic?

With respect to the Fermi adjoint
$\dag$, half of the Clifford generators (the $\gamma_i$)
are Hermitian and the other half (the $\tilde{\gamma}_i$)
are anti-Hermitian.
In a Clifford $\dag$-algebra, however,
all the generators are anti-Hermitian or Hermitian
together.
Therefore the  Clifford-algebra  generators
$\{\gamma_i, \, \tilde{\gamma}_i\}_{i=1}^n$
are not Clifford $\dag$-algebra generators.

In some cases we construct suitable generators
using the top element $\tilde{\gamma}^{\uparrow}$ of
$\tilde{E}$:

If the dimension $n$ of $E$ (and $F$) is a
multiple of 4, then
$\bar{\gamma}_i:=\tilde{\gamma}^{\uparrow}
\tilde{\gamma}_i$ anticommutes with the $\gamma_j$,
and is Hermitian like the $\gamma_j$.
Then the elements $\{\gamma_i, \, \bar{\gamma}_i\}_{i=1}^n$
generate a Clifford  $\dag$-algebra with ({\it cf.}
(\ref{eq:CLIFFORDW}))
\BEqA
   \gamma_i\gamma_j+\gamma_j\gamma_i=+2\delta_{ij}, \nonumber
\\ \bar{\gamma}_i\bar{\gamma}_j
+ \bar{\gamma}_j\bar{\gamma}_i=+2\delta_{ij},
\nonumber \\
\bar{\gamma}_i \gamma_j+\gamma_j\bar{\gamma}_i=0,
\EEqA
which is isomorphic to the F-D algebra of
$F$.
Then the Clifford-quantified $\dag$-algebra (the case
$\zeta=\zeta' = +1$) is isomorphic to a
Fermi-quantified one when
$n=4m$, and the adjoint of the one-cliffordon space is
positive definite.
The two quantified theories then predict the
same transition amplitudes and spectra.

Analogously, when $\zeta =\zeta' = -1$
and all the Clifford generators are
anti-Hermitian, and $n=4m$, the F-D and
Clifford statistics again give isomrphic $\dag$-algebras.

They still differ in their grades.
The F-D quantified
system has a grade $G_F$ with spectrum $-N, \dots, 0, \dots, N$,
corresponding to the creation and annihilation fermions.
     The Clifford quantified system
has a positive grade operator $G_C$ with
spectrum $0, 1, \dots 2N$.
The operators $G_C$ and $G_F$ do not even commute.
The F-D and Clifford graded-algebras are not isomorphic.

This is merely a difference in language  The operators
that are said to create and annihilate things in F-D
statistics are said to permute things in Clifford
statistics. In Clifford statistics nothing is created or
destroyed.

\section*{ACKNOWLEDGMENTS}
This work was aided by discussions with James Baugh.
It was partially supported by the M. and H. Ferst
Foundation.

\end{document}